\documentstyle[prc,aps,psfig]{revtex}
\newcommand{\MeV}{\:{\rm MeV}}
\newcommand{\keV}{\:{\rm keV}}
\newcommand{\fm}{\:{\rm fm}}

\newcommand{\be}{\begin{equation}}
\newcommand{\ee}{\end{equation}}
\newcommand{\bea}{\begin{eqnarray}}
\newcommand{\eea}{\end{eqnarray}}
\begin{document}
\draft
\title{Isospin singlet ($\boldmath pn$) pairing and quartetting contribution to
  the binding energy of nuclei} 
\author{G.\ R\"opke, A.\ Schnell, P.\ Schuck\cite{isn}}
\address{University of Rostock, FB Physik,
Universit\"atsplatz 1, 18051 Rostock, Germany}
\author{ U.\ Lombardo}
\address{Dipartimento di Fisica, 57 Corso Italia, I-95129 Catania and\\
  INFN-LNS, 44 Via S.~Sofia, I-95123 Catania, Italy}
\date{\today}
\maketitle
\begin{abstract}  
  Isospin singlet ($pn$) pairing as well as quartetting in nuclei is
  expected to arise near the symmetry line $N=Z$. Empirical values
  can be deduced from the nuclear binding energies applying special
  filters. Within the local density approximation, theoretical
  estimates for finite nuclei are obtained from results for the
  condensation energy of asymmetric nuclear matter. It is shown that
  the isospin singlet condensation energy drops down abruptly for
  $|N-Z|\approx 4$ for medium nuclei in the region $A=40$. Furthermore,
  $\alpha$-like quartetting and the influence
  of excitations are discussed.

\end{abstract}


\section{Introduction}
The structure of nuclei is understood to a large extent within a
single-particle approach, the nuclear shell model which was
introduced in nuclear theory 50 years ago \cite{shell-model}. Its microscopic
foundation should be given from a quantum statistical treatment of the 
many-nucleon system, where in general also the concept of temperature
can be introduced to describe excited nuclei in a dense medium. Within 
the mean-field approximation, the single-particle approach can be
obtained \cite{BM}.

However, due to the nucleon-nucleon interaction also correlations
occur which cannot be accounted for within this quasi-particle
approach. Two-particle correlations and higher order correlations as
precursors of corresponding bound states may become of importance
in regions where the nucleon density is low. A systematic treatment of 
the effect of correlations can be given in terms of the spectral
function, see, e.g., \cite{Arne,Wim} and further references given therein.

Free two-nucleon states are described by scattering phase shifts or,
more generally, by the $T$ matrix. In the isospin singlet ($S=1,\,T=0$)
channel where the interaction is stronger than in the isospin triplet
($S=0,\,T=1$) channel, a bound state, the deuteron, is formed.
Two-nucleon states in nuclear matter are strongly modified when increasing
the density. In particular, the bound
state will disappear at the so-called Mott density \cite{RMS82,SRS}.

Another aspect of the nucleon-nucleon interaction is the formation of
quantum condensates. At low temperatures it is well known that in
nuclei, nuclear matter, and neutron matter (neutron stars) superfluidity
can arise in the isospin triplet channel \cite{BM,Nazarew,ShapT}.
Nucleonic pairing is a well-known effect in the structure of nuclei.
In the isospin triplet channel the influence of
proton-proton ($pp$) or neutron-neutron ($nn$) pairing on the binding
energy of nuclei has extensively been investigated, cf.~\cite{BM}.

However, under certain conditions the interaction in the
proton-neutron ($pn$) isospin singlet channel may be even
stronger. For instance, in the two-particle system a bound state, the
deuteron, arises. In nuclear matter, at densities below the Mott
density a bound state (quasi-deuteron) can exist so that at low
temperature the Bose-Einstein condensation of these quasi-deuterons
may occur.  An interesting feature of the isospin singlet
pairing in symmetric nuclear matter is the cross-over from
Bose-Einstein condensation of deuterons at low densities to BCS
neutron-proton pairing at high densities \cite{BLS,ZPhys95}.

In spite of the relatively strong interaction,  
triplet pairing seems less apparent in nuclear structure
systematics (see the studies of Goodman on $T=0$ pairing in nuclei
\cite{goodman}). However, it should be essential for nuclei near the
symmetry line $N\approx Z$. This phenomenon becomes
of importance for heavier $N\approx Z$ nuclei \cite{goodman2} as will be
produced in the new radioactive beam facilities.

A further interesting effect is the possible occurrence of higher
order condensates such as $\alpha$-like quartetting \cite{Marumori,Cau81}.
At present, the signatures of isospin singlet pairing and the relation to
quartetting have not so clearly been worked out.  An interesting point
is to identify signatures of quartetting in finite nuclei, see
\cite{lett}. One of the possible consequences is the contribution of
isospin singlet pairing and four-particle correlations to the
binding energy of nuclei, that will be discussed below.

The treatment of correlations in the many-nucleon system is more
simple for infinite nuclear matter. The new results with
respect to isospin singlet pairing and quartetting \cite{lett}
should be of interest for the evaluation of binding
energies of finite nuclei as well. One possible way to apply results
of infinite matter calculations to finite systems is the local density
approximation (LDA) which was elaborated successfully in atomic,
molecular and condensed matter systems, see \cite{LDA} and references
therein quoted. We will use this approach to perform exploratory
calculations in finite nuclei. Compared with shell-model calculations,
the LDA is by far more simple and has been proved to give adequate
results in the case of $pp$ and $nn$ pairing \cite{ST}.  We will use
this approach to give estimates for the effects of isospin singlet
pairing and quartetting on nuclear binding energies.

The question of different kinds of
pairing in proton-rich nuclei has been investigated by various
authors. In addition to the $T=1$ pairing, $T=0$ $np$-pairing
modes have been investigated to explain the Wigner energy in $N=Z$
nuclei \cite{Satula97}. A more extended discussion of the origin of
the Wigner term is given in \cite{Satula97a} where a method has been developed
to extract the Wigner term from experimental data. Both, empirical
arguments and shell-model calculations suggest that the Wigner term
can be traced back to the isospin $T=0$ part of nuclear
interaction. However, it has also been found
that the Wigner term cannot be solely
explained in terms of correlations between the neutron-proton $J=1,\,T=0$
(deuteron-like) pairs. One should, however, realize that in a
finite nucleus the intrinsic quantum numbers of the deuteron can be
mixed with the orbital motion of the nucleons in the shell model potential.
Recently, pairing and the structure of the $pf$-shell $N \approx Z$
nuclei have been discussed in \cite{Poves98}. There, the isovector and
isoscalar pairing interaction has been studied. It has been found that the
Wigner energy cannot purely be explained as a pairing effect,
considering only the zero angular momentum ($L=0$) channel.

Calculations of the pairing and isospin symmetry in proton-rich nuclei 
were performed \cite{Engellanganke}. Near $N\approx Z$, a steep
decrease of the isoscalar proton-neutron pairing energy is found with
increasing $|N-Z|$.
An interesting fact is that in $^{50}$Cr proton-proton and neutron-neutron
pairing is more reduced with temperature than the isoscalar
proton-neutron pairing \cite{Langanke}.

Proton-neutron pairing has been discussed also with respect to other
properties, see \cite{lett}. For instance, a cranked shell model for
the description of rotational bands in $N \approx Z$ nuclei has been
formulated in \cite{Frauendorf} taking the isovector proton-neutron
pairing explicitly into account. More detailed discussions with
respect to the binding energies can be found in \cite{Cau81}.
Proton-neutron pairing and its relation to isospin symmetry has been studied
within the BCS theory in \cite{Civitarese}.
Recently, the role of the proton-neutron interaction in $N\approx Z$ nuclei
and its consequences for $pn$-pairing has been investigated in \cite{Kaneko}.

We will proceed as follows: In Section 2 we analyze the binding 
energies of nuclei near $N=Z$. The basic formalism of LDA
is given in Section 3. We proceed then with the presentation of
numerical results in Section 4 and draw some conclusions in Section 5.
\section{Binding energies of nuclei near $N=Z$}
\subsection{Filters for isospin singlet condensates}
The empirical binding energies of nuclei are well approximated by the
Bethe-Weizs\"acker formula, containing the contributions of bulk,
surface, Coulomb and asymmetry energy. Additional contributions
originate from pairing in the isospin triplet channel ($pp$, $nn$) and
shell structure effects, for instance the behavior near magic numbers.

We assume that the binding energy $B(Z,N)$ can be decomposed into
different contributions as
\begin{eqnarray}
\label{BW}
B(Z,N)&=& B_{\rm bulk}(Z,N) +  B_{\rm surf}(Z,N) + B_{\rm Coul}(Z,N) +
B_{\rm asymm}(Z,N) \nonumber\\
& +& B_{\rm shell}(Z,N) + B_{\rm pair}(Z,N) + \Delta B(Z,N)\,\,. 
\end{eqnarray}
The term $\Delta B(Z,N)$ contains the effects of correlations between
protons and neutrons (isospin singlet ($pn$) pairing, quartetting) which
are of interest here. In nuclear matter, the occurrence of different
condensates has been intensively investigated.
Isospin singlet ($pn$) pairing is
strong in symmetric nuclear matter for densities up to the saturation
density. The dependence of the $pn$ pairing gap on the
asymmetry of nuclear matter for different densities has been
investigated in \cite{AFRS,SAL}. If the densities of protons and
neutrons, or the corresponding chemical potentials $\mu_p$, $\mu_n$
are sufficiently different, a condensate cannot be formed.

Analogously, the effects of isospin singlet ($pn$) pairing and
quartetting are expected to be strong near  $N=Z$, and disappear
if the absolute value of the difference $N-Z$ becomes large. 
Therefore, it seems to be reasonable to investigate the behavior of  
\begin{equation}
\Delta B(Z,N)=b_{N-Z}(Z)
\end{equation}
in dependence of $N-Z$, considering $Z$ as a parameter.   

To select out the contribution $\Delta B(Z,N)$ to the binding energy,
filters can be applied which eliminate the other contributions given
in (\ref{BW}) to a large extent. We consider the following (horizontal 
and vertical) filters
\begin{eqnarray}
h(Z,N)&=&2 B(Z,N)- B(Z,N-2) -B(Z,N+2)\nonumber\\
&&-{}2B(Z-2,N)+ B(Z-2,N-2)+ B(Z-2,N+2)\,,\label{h}
\end{eqnarray}
\begin{eqnarray}
v(Z,N)&=&2 B(Z,N)- B(Z-2,N) -B(Z+2,N)\nonumber\\
&&-{}2B(Z,N+2)+ B(Z-2,N+2)+ B(Z+2,N+2)\,.\label{v}
\end{eqnarray}
These simple filters contain the difference of second order
differences. Therefore, smooth dependences on $Z$, $N$ nearly cancel
out, up to second order. This concerns not only the contributions of bulk,
surface, and Coulomb energy, but also the asymmetry term. The isospin
triplet pairing effects compensate because only nuclei of equal
parity are considered. Similarly, shell effects compensate to a
certain extent assuming that they are related only to the respective
values of the numbers $Z$ or $N$.

From these primitive filters (\ref{h}), (\ref{v}), other filters can
be constructed by superposition. 
The filter used in \cite{Satula97} to evaluate the Wigner energy can
be constructed as 
\begin{equation}
\label{sat}
W(A) = -{1 \over 8}\,\, v\left({A \over 2}, {A \over 2}-2\right) -{1
  \over 8}\,\, h\left({A 
  \over 2}, {A \over 2}-2\right)\,\,,
\end{equation}
or, more generally
\begin{equation}
\label{w}
w(Z,N) = -\frac{1}{8}\,\, v(Z, N-2) -\frac{1}{8}\,\,
h(Z, N-2)\,.
\end{equation}
Another, more symmetric filter
\begin{equation}
\label{g}
g(Z,N) = {1 \over 8} h(Z, N) -{1 \over 8} h(Z+2, N )
\end{equation}
was considered in \cite{RSSnashville}.

\begin{figure}
\centerline{\psfig{figure=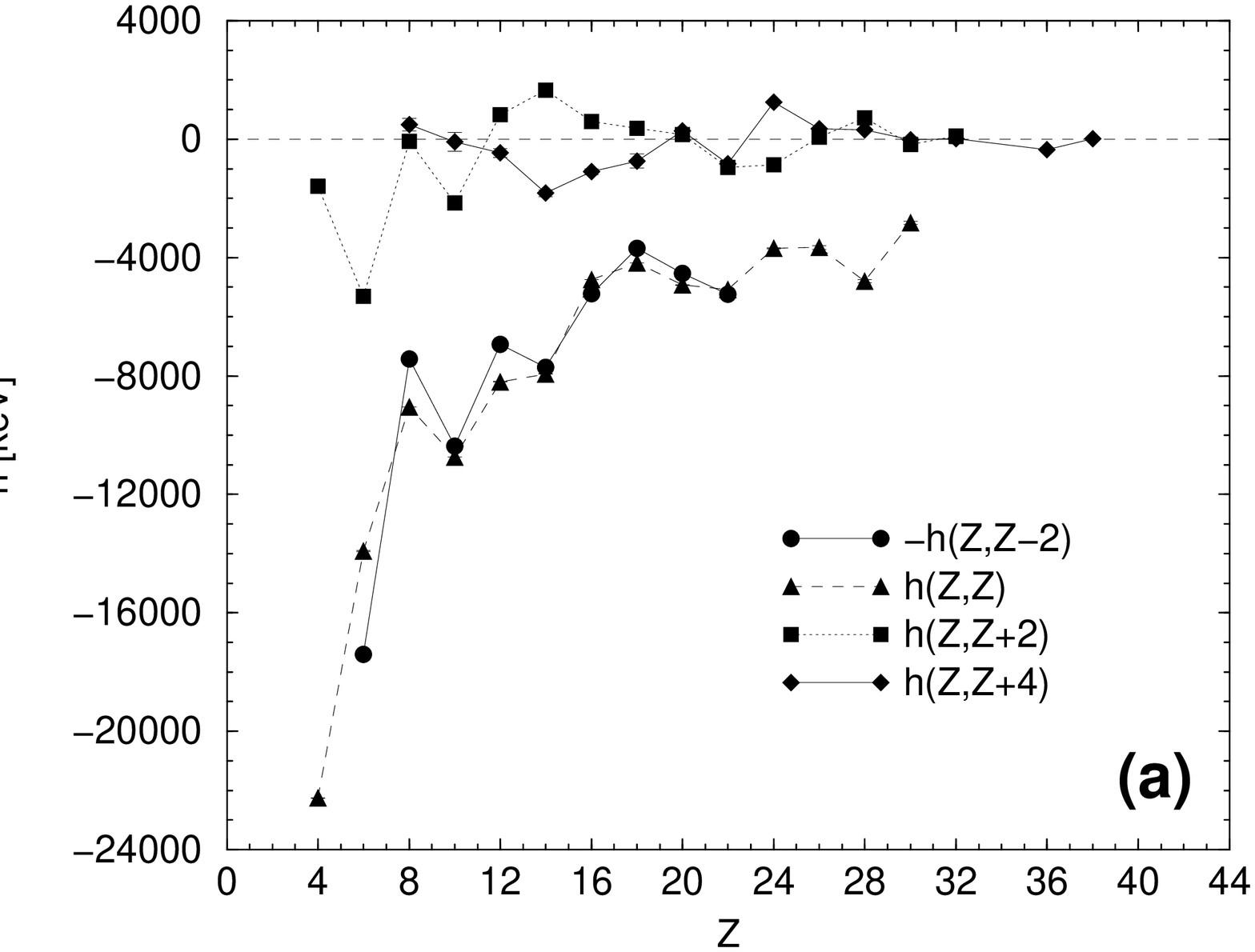,width=8cm}}
\centerline{\psfig{figure=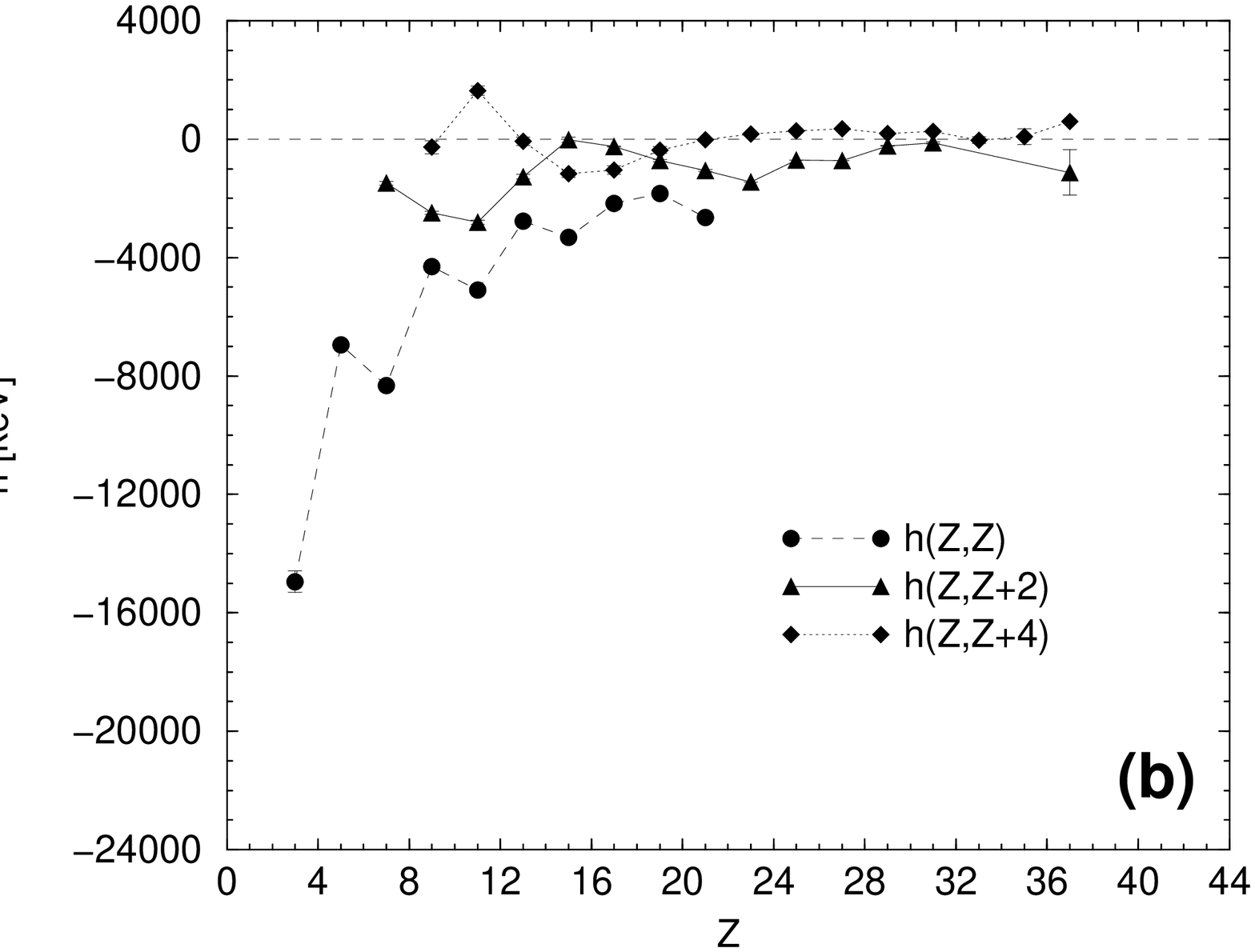,width=8cm}}
\centerline{\psfig{figure=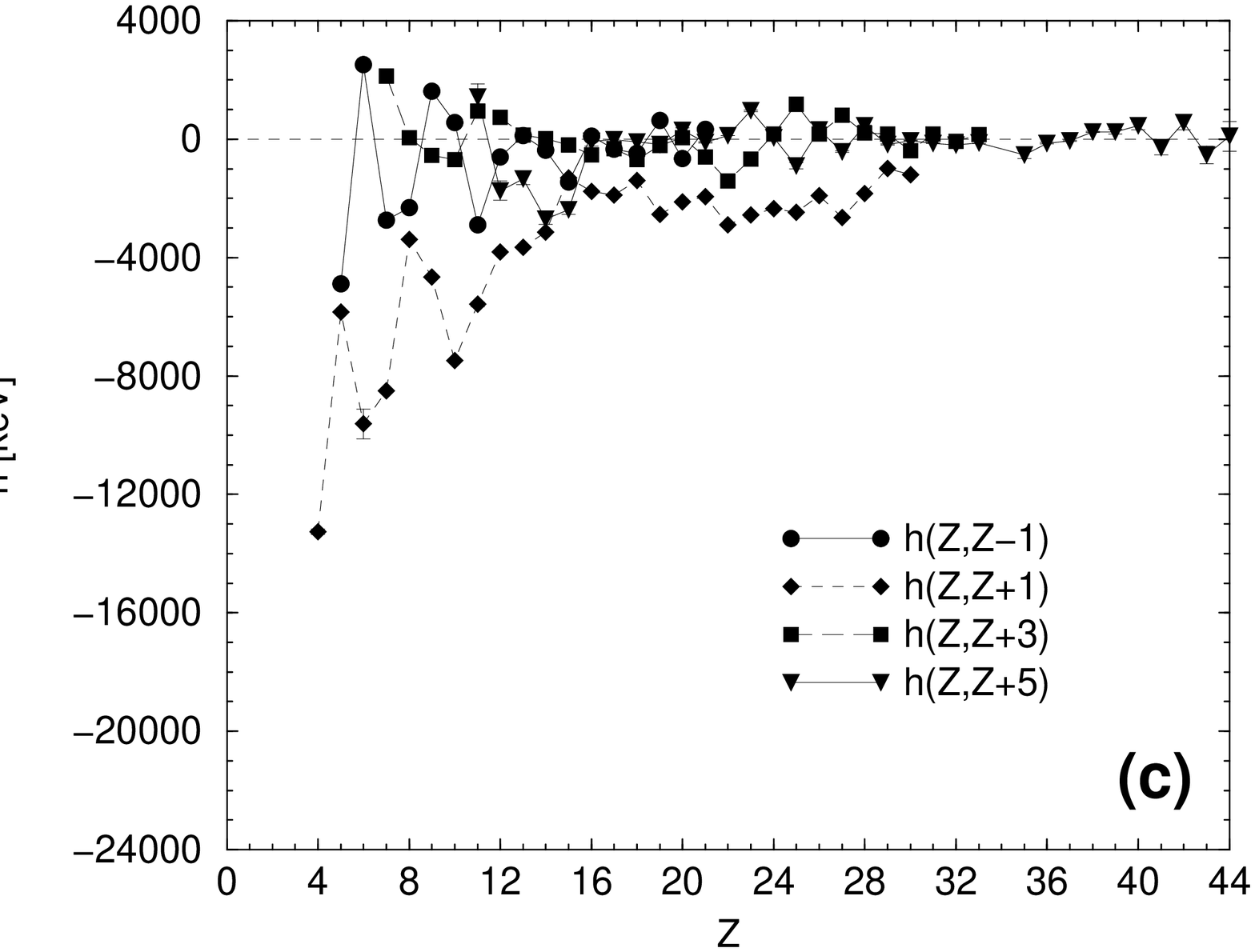,width=8cm}}
\caption{The filter $h(Z,Z+i)$, Eq. (\ref{h}), is given for various $i$
as it can be extracted from the experimental binding energies.
Results are separately shown for even-even (a), odd-odd (b),
and even-odd/odd-even (c) nuclei.}
\label{fig_h-filter}
\end{figure}

We give the results for the filter $h(Z,N)$ in Fig.~\ref{fig_h-filter}.
The dependence
of $h(Z,N)$ on $Z$ is shown for different parameter values $N-Z$,
where even and odd values of $Z$ are shown separately. It is clearly
seen that the filter $h(Z,N)$ becomes small for large
$N-Z$. Furthermore, the average value decreases with increasing
$Z$. Also, even-even and odd-odd numbers of $Z,N$ show a different
behavior. Similar results can be obtained using the filters (\ref{v}), (\ref{w}), or (\ref{g}).

The relation to the quantities $b_i(Z)$ introduced above can
immediately be given. For the filter $h(Z,N)$ we find
\begin{eqnarray}
h(Z,N)&=&-b_{N-Z-2}(Z)+2b_{N-Z}(Z)+b_{N-Z}(Z-2)\nonumber\\
&&-{}b_{N-Z+2}(Z)-2b_{N-Z+2}(Z-2)+b_{N-Z+4}(Z-2)\,.
\end{eqnarray}
Since only differences of the quantities $b_i(Z)$ are determined by
the filter $h(Z,N)$, their values can be reconstructed using
additional assumptions. To simplify the extraction of the quantities
$b_i(Z)$  we assume that $b_{Z-N}(Z)$ is strongly depending on the
asymmetry $(Z-N)$, but only weakly depending on the mass number $(Z)$, 
after eliminating a possible even-odd staggering.
Therefore, we approximate $b_i(Z-2)
\approx b_i(Z)$ so that $h(Z,N) \approx -b_{N-Z-2}(Z)+3b_{N-Z}(Z) -
3b_{N-Z+2}(Z) +b_{N-Z+4}(Z)$.
According to the increase of asymmetry discussed above, we assume that
the $b_i(Z)$ disappear for large absolute values of $i$. Taking $b_i=0$
for $|i|\geq 6$, the results for the quantities 
$b_i(Z)$ are shown in Fig.~\ref{fig_bh-absolut}.

\begin{figure}
\centerline{\psfig{figure=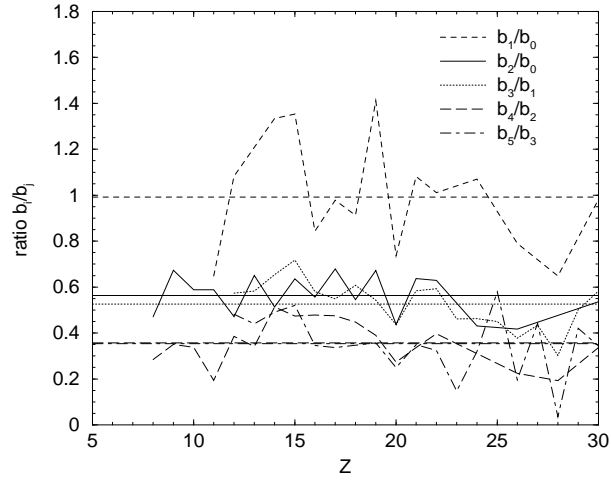,width=8cm}}
\caption{A selection of several ratios $b_i/b_j$ as a function of $Z$.
The dashed lines represent the average values of the ratios.}
\label{fig_bh-ratio}
\end{figure}

\begin{figure}
\centerline{\psfig{figure=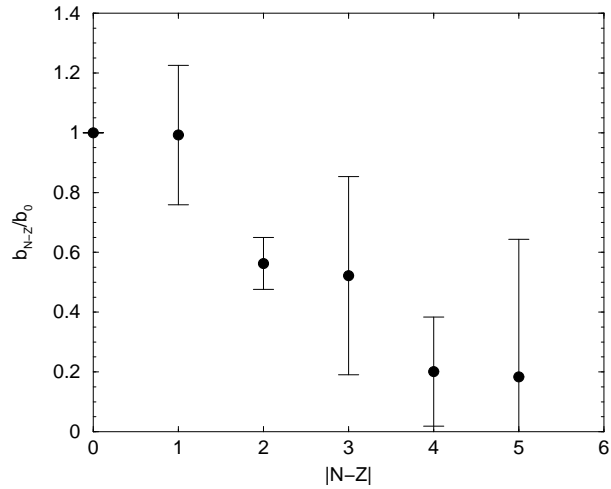,width=8cm}}
\caption{Averages of the ratio $b_{N-Z}/b_0$ as a function of the absolute
value of the difference between protons and neutrons.}
\label{fig_bh-diff}
\end{figure}

A first important result is that the quantities $b_i(Z)$ depend mainly 
on the absolute value of $i$. As a consequence, $h(Z,Z-2)$ nearly
coincides with $-h(Z,Z)$, whereas $h(Z,Z-1)$ is close to zero, in
contrast to $h(Z,Z+1)$.
In detail, the values for $b_0$, $b_2$,
$b_4$ were found from the solution of equations containing
$h(Z,Z)$, $h(Z,Z+2)$, $h(Z,Z+4)$, whereas the values for $b_1$, $b_3$,
$b_5$ were found from the solution of equations containing
$h(Z,Z+1)$, $h(Z,Z+3)$, $h(Z,Z+5)$. To derive the average
behavior with respect to $Z$, we considered the ratios $b_i/b_0$ as
shown in Fig.~\ref{fig_bh-ratio}.
The average of these values over the whole range of experimentally
accessible data ($8\le Z\le 30$) are given in Fig.~\ref{fig_bh-diff}.
They show a
decrease with increasing $|i|=|N-Z|$, so that $b_5$ is nearly zero.
\subsection{Further filters}
\begin{figure}
\centerline{\psfig{figure=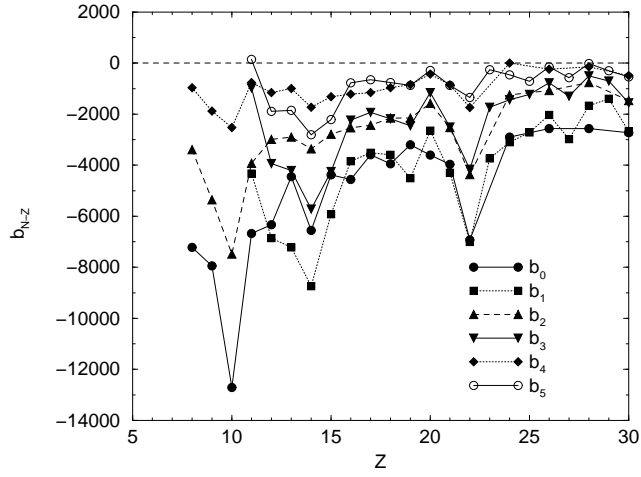,width=8cm}}
\caption{The parameter $b_{N-Z}(Z)$ for even and odd proton number $Z$
  as it is derived from the filter $h$.} 
\label{fig_bh-absolut}
\end{figure}

\begin{figure}
\centerline{\psfig{figure=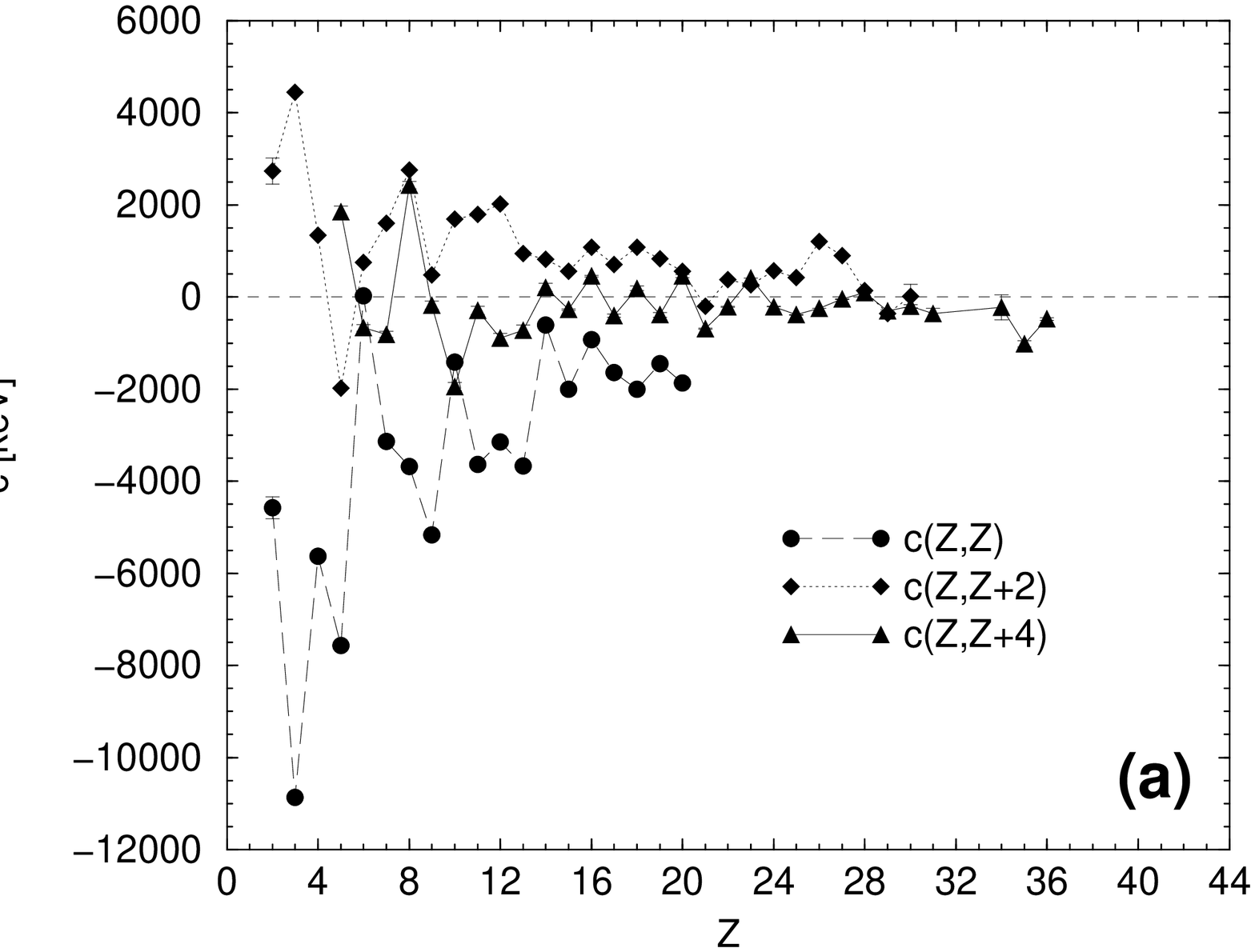,width=8cm}}
\centerline{\psfig{figure=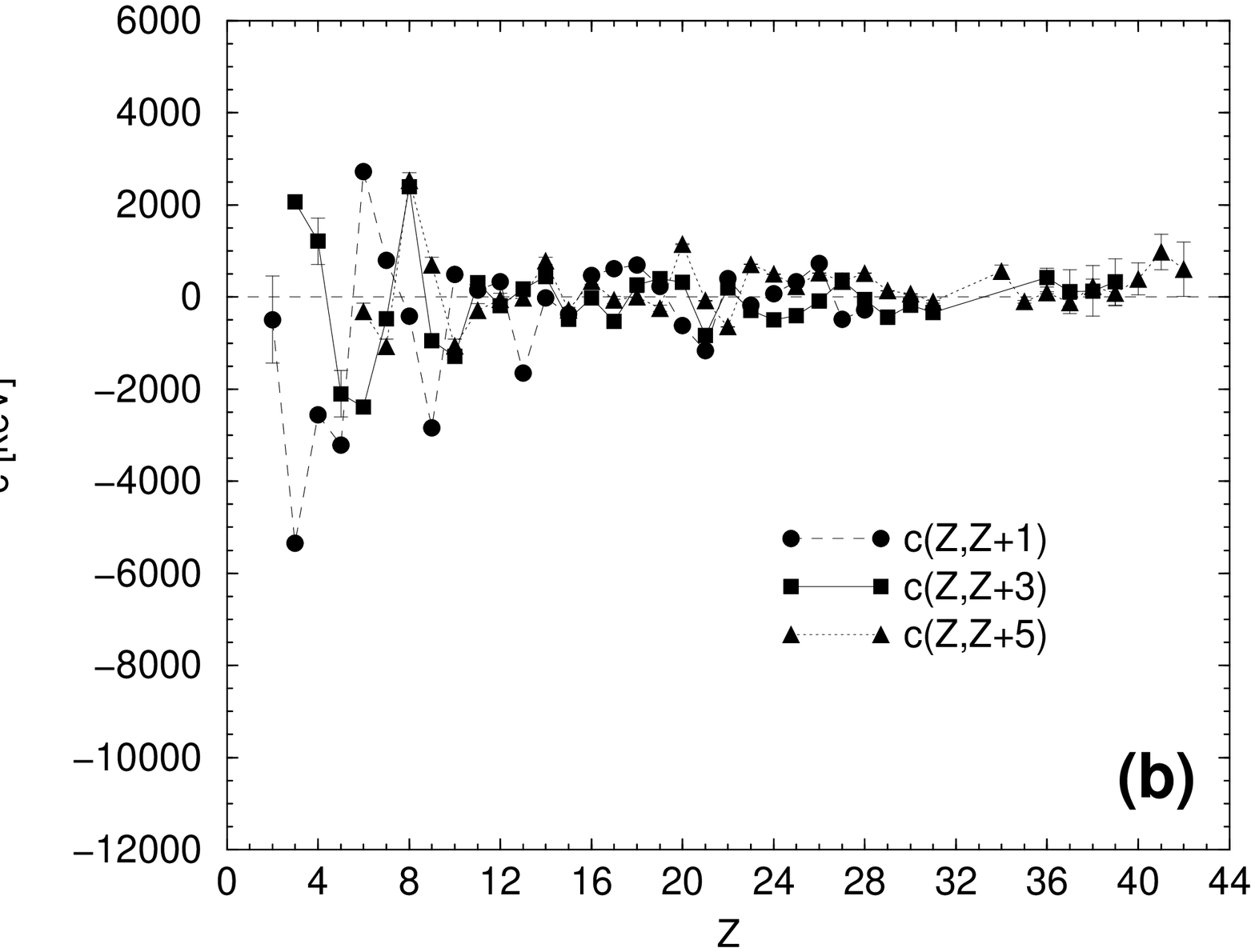,width=8cm}}
\caption{The filter $c(Z,Z+i)$ for various $i$.
Results are separately shown for even-even/odd-odd (a)
and even-odd/odd-even (b) nuclei.}
\label{fig_c-filter}
\end{figure}

\begin{figure}
\centerline{\psfig{figure=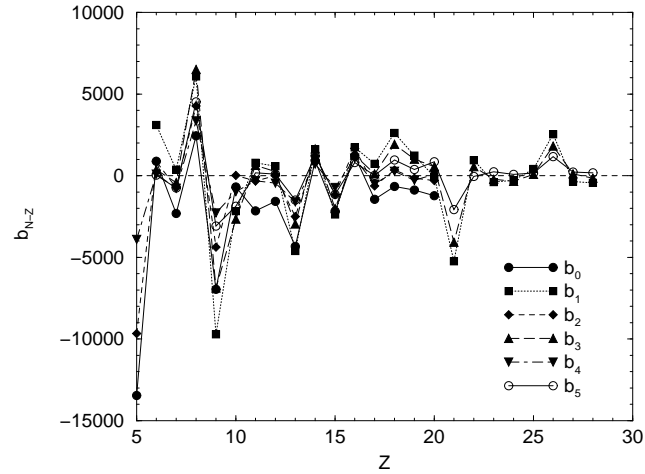,width=8cm}}
\caption{The parameter $\delta b_{N-Z}(Z)$ as derived from the filter
  $c$, Eq. (\ref{c}), for
even and odd proton number $Z$.}
\label{fig_bc-absolut}
\end{figure}

\begin{figure}
\centerline{\psfig{figure=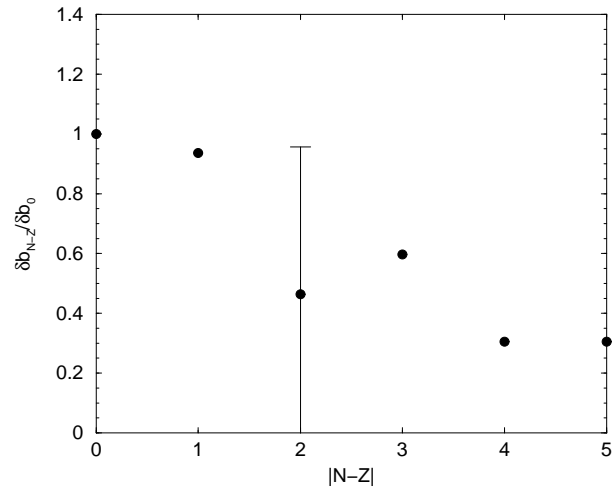,width=8cm}}
\caption{Averages of the ratio $\delta b_{N-Z}/\delta b_0$ as a function of the
absolute value of the proton-neutron difference.}
\label{fig_bc-diff}
\end{figure}

A further interesting property of the quantities $b_i(Z)$ is their
dependence on $Z$. First we are interested in a possible even-odd
staggering. This can be deduced from the values shown in
Fig.~\ref{fig_bh-absolut}.
Alternatively, we also can consider, as a special indicator to
extract even-odd staggering, the filter
\begin{eqnarray}
\label{c}
c(Z,N)&=&(-1)^Z \left[ B(Z,N)- B(Z,N+2) -B(Z+2,N)+ B(Z+2,N+2)
\right.\nonumber\\ 
&+&\left.+B(Z+1,N-1)- B(Z+1,N+1) - B(Z-1,N-1)+ B(Z-1,N+1)\right]\,.
\end{eqnarray}
We decompose
\begin{equation}
\label{bi_even-odd}
b_i(Z) = \bar b_i(Z) + (-1)^Z \delta b_i(Z) 
\end{equation}
and assume, that the remaining dependence on $Z$ is smooth, as already 
used above. Then we have
\begin{equation}
c(Z,N)=-2 \delta b_{N-Z-2}(Z) + 4  \delta b_{N-Z}(Z) -2
 \delta b_{N-Z+2}(Z)\,\,. 
\end{equation}
Now, we can perform the reconstruction of $\delta b_i(Z)$ along
the lines given for $b_i(Z)$. The filter $c(Z,N)$ is shown in
Fig.~\ref{fig_c-filter},
the parameters $\delta b_i(z)$ in Fig.~\ref{fig_bc-absolut} as a function of
$Z$, and the averaged values are shown in Fig.~\ref{fig_bc-diff}
vs the asymmetry $i=N-Z$. We see that also there the staggering contribution
disappears with increasing $|i|$, however, the statistical errors are
large. The global dependence of $\bar b_i(Z),\,\,\delta b_i(Z)$ on $Z$
cannot easily be extracted. After a steep decrease for small $Z$ a
flattening is observed at higher values of $Z$,
see also Figs.~\ref{fig_bh-absolut}, \ref{fig_bc-absolut}.

\begin{figure}
\centerline{\psfig{figure=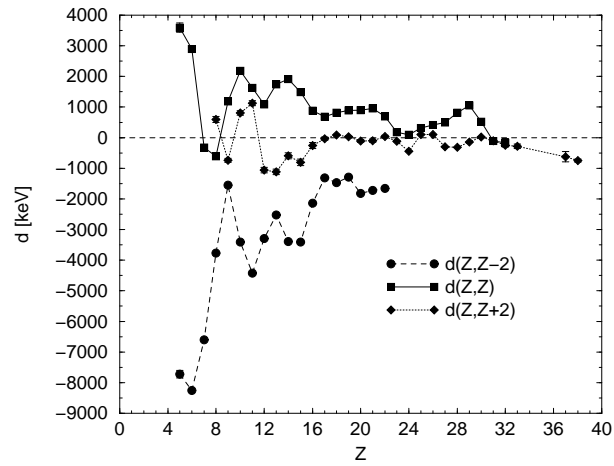,width=8cm}}
\caption{For demonstration purposes the filter $d(Z,Z+i)$ for various
$i$ is given considering only even-even and odd-odd nuclei.}
\label{fig_d-filter}
\end{figure}

From other approaches, see \cite{Satula97a}, the Wigner energy $E_W$
is introduced which 
occurs in the additional binding due to the $np$-pair correlations 
\begin{equation}
B_{np, {\rm pair}} = - \epsilon_{np}(A) \pi_{np} +E_W
\end{equation}
with $\pi_{np}=(1-(-1)^N)(1-(-1)^Z)/4$. The first contribution to the
$np$ pairing energy ($\epsilon_{np}$) represents additional binding 
due to the residual interaction between the two odd nucleons in an
odd-odd nucleus. The Wigner energy $E_W$ is believed to represent the
energy of collective $np$-pairing correlations. It can be decomposed
into two parts:
\begin{equation}
E_W=W(A) |N-Z|+d(A)\pi_{np} \delta_{NZ}
\end{equation}
Filter used in \cite{Satula97} can be constructed as given by
(\ref{sat}) and
\begin{equation}
d(A) = {1 \over 2}\,\, c\left({A \over 2}-3, {A \over 2}-1\right) +{1
  \over 2}\,\, c\left({A 
  \over 2}-2, {A \over 2}\right)\,\,. 
\end{equation}
The $d$-term represents a correction for $N=Z$ odd-odd
nuclei. Estimates suggest that the ratio $d/W$ is constant, values 1
and 0.56 have been reported \cite{Satula97}.
Both filters can be expressed by superposition in terms of 
the more simple filters $h(Z,N),\,\,v(Z,N),\,\,c(Z,N)$ given here.
A more general filter is introduced as
$d(Z,N)=c(Z-3,N-1)/2 +c(Z-2,Z)$ as an average over neighboring values
of $c$. Some values are shown in Fig.~\ref{fig_d-filter}. 

The results of our phenomenological treatment can be summarized as
follows (some preliminary results have been presented in \cite{erice}): 
\begin{itemize}
\item[(i)]
There is a contribution $b_i(Z)$ to the nucleon binding
energy which is due to proton-neutron correlations of the type of a
isospin-singlet pairing or a quartetting condensate.
\item[(ii)]
This contribution depends on the absolute value of the asymmetry
parameter $|N-Z|$. It has a maximum magnitude for symmetric nuclei
$N=Z$ and decreases with increasing asymmetry, disappearing near
$|N-Z|=4$.
\item[(iii)]
It shows an even-odd staggering as function of $Z$.
\item[(iv)]
On the average, it decreases with increasing $Z$, steep for small 
values of $Z$, but flat for large $Z$.
\end{itemize}
\section{Local Density Approximation}
A theoretical interpretation of the contribution to the
binding energy due to $pn$ pairing and quartetting could be given by
the Local Density Approximation (LDA). In contrast to
shell-model calculations, the LDA is by far more simple and
has been proved to give adequate results in the case of $pp$ and $nn$
pairing \cite{ST}.

As well known from quantum statistics of the inhomogeneous fermion
gas, the energy and wave function of the ground state of a
many-fermion system can be calculated within a variational
approach. The energy density is considered as a functional of the
fermion density, which, in the case of a nucleonic system, depends in
addition to space coordinates also on spin and isospin. It can be
decomposed into kinetic, potential, exchange and correlation energy. Within a
gradient expansion, in lowest order the energy density depends only on 
the local values of the nucleon density. As a consequence, the
exchange and correlation energy can be approximated using results from 
nuclear matter calculations.

In particular, the contribution to the exchange and correlation part
of the energy due to the formation of a condensate can be evaluated
within nuclear matter theory. There is an extended literature on
isospin triplet ($pp,\,nn$) pairing. More recently, also isospin
singlet ($pn$) pairing has been considered \cite{BLS}, which for
symmetric nuclear matter may become stronger compared with
isospin triplet pairing at subnuclear densities because of the
more attractive nucleon-nucleon interaction in the isospin singlet channel.
\subsection{Pairing vs. quartetting in symmetric matter}
A standard way to describe quantum condensates in many-body systems is 
the method of thermodynamic Green functions.
Treating the two-particle Green function in
ladder Hartree-Fock approximation, an effective wave equation (in matrix
notation) $\psi_{\lambda}=K_2(E_{\lambda})\,\psi_{\lambda}$ for the quantum
state $\lambda$ can be derived. Explicitly this reads
\be
\psi_{\lambda}(12) = \sum_{1'2'} K_2(12,1'2', \epsilon_{\lambda})
\,\psi_{\lambda}(1'2')  
\label{two_wave}
\ee
with
\be
K_2(12,1'2',z) = V(12,1'2')\frac{1-f(1)-f(2)}{z-\epsilon(1)-\epsilon(2)}\:.
\label{two_kernel}
\ee
The influence of the medium is contained in the
single-particle energy
\begin{equation}
\label{eps}
\epsilon(1)=p_1^2/2m+\sum_{2}V(12,12)f(2)
\end{equation}
and in the Pauli blocking term $[1-f(1)-f(2)]$.
Here, $f(1)=[\exp\{\epsilon(1)/T-\mu/T\}+1]^{-1}$ is the Fermi distribution
function and '1' denotes momentum, spin, and isospin coordinates,
whereas $V(12,1'2')$ is the antisymmetrized matrix element of the
two-body interaction.

The transition to a superfluid state is obtained from the Thouless
criterion 
as described by the Gorkov equation $\psi_2=K_2(\mu_1+\mu_2)\,\psi_2$.
Depending on the respective channels considered, it allows the
determination of the critical temperatures $T_s^c$ or $T_t^c$ for the
isospin singlet and triplet channels, respectively,
as a function of the chemical potential.

\begin{figure}
\centerline{\psfig{figure=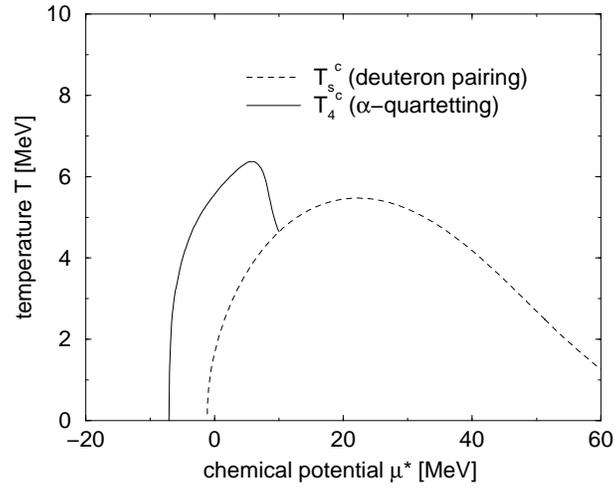,width=8cm}}
\caption{Critical temperatures for the onset of quantum condensation in
symmetric nuclear matter, model calculation. The critical temperature of the
onset of two-particle pairing $T_t^c$ is compared with $T_4^c$ for the onset
of a four-particle condensate, as a function of the chemical potential.}
\label{fig_tc-mue}
\end{figure}

\begin{figure}
\centerline{\psfig{figure=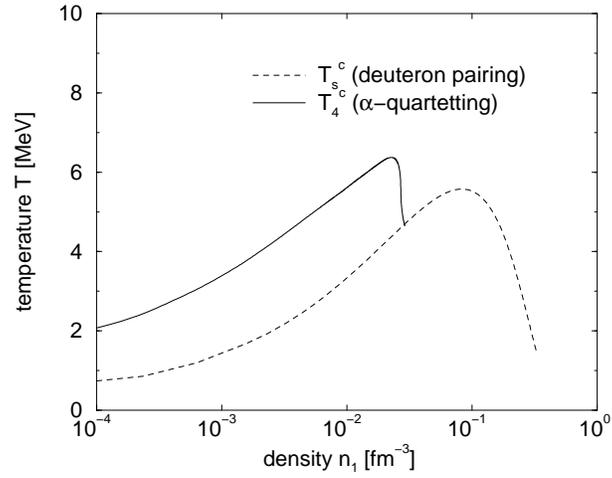,width=8cm}}
\caption{The same as Fig.~\protect\ref{fig_tc-mue} but as a function of the
uncorrelated density.}
\label{fig_tc-n}
\end{figure}

The solution of the Gorkov equation has been considered by different 
authors using realistic bare nucleon-nucleon interactions. It has been found
that in comparison  with the isospin triplet channel, in the isospin
singlet channel 
the transition to superfluidity  should arise at relatively high
temperatures \cite{BLS,AFRS,SAL},
see also Figs.~\ref{fig_tc-mue}, \ref{fig_tc-n}.
This is a consequence of the stronger 
interaction in the isospin singlet channel which leads to the formation of
the deuteron in the low-density limit where $f\ll 1$. 
Estimates give a value of the critical temperature up to
$T_s^c\approx 5\MeV$ at one third of the nuclear matter density.
At the same time, at zero temperature a large gap arises \cite{BLS}.

In a recent letter \cite{lett} it has been shown that in a certain
region of density, pairing has to compete with quartetting. It has
been found that under certain conditions in symmetric nuclear matter
the transition to isospin 
singlet pairing, which is stronger than triplet pairing, will not occur
because the quartetting transition occurs before that. Within a 
cluster-mean field approach \cite{RMS82,dukelski}, the
critical temperature for the quartetting transition was obtained from
the equation
\begin{eqnarray}\label{g4_full}
G_4(1234,1'2'3'4',z) & = & \frac{f(1)f(2)f(3)f(4)}{g_4(1234)}
\,\frac{\delta_{11'}\delta_{22'}\delta_{33'}\delta_{44'}}
{z-\epsilon_4(1234)} \nonumber\\
&&+{}\sum_{1''2''3''4''}\!\!\!\!\!K_4(1234,1''2''3''4'',z)
G_4(1''2''3''4'',1'2'3'4',z)\,,\!\!
\end{eqnarray}
\begin{eqnarray}\label{k4_full}
K_4(1234,1'2'3'4',z) &=& V(12,1'2')
\frac{f(1)f(2)}{g_2(12)} \frac{\delta_{33'}\delta_{44'}}{z-\epsilon_4(1234)}
+\,{\rm perm.}\,,
\end{eqnarray}
where we use the abbreviation $\epsilon_n(12\dots n)=\epsilon(1)+\epsilon(2)+\cdots+\epsilon(n)$, and
$g_n(12\dots n) = [\exp(\epsilon_n(12\dots n)-n\mu)/T-1]^{-1}$ being the Bose
distribution function.
The instantaneous part of interaction kernel is obtained by using the
technique of Matsubara Green functions as
where the terms obtained by renumbering are not given explicitly.
We have used the identity $\bar{f}(1) \bar{f}(2) \cdots \bar{f}(n)
-f(1) f(2) \cdots f(n) = g_n^{-1}(12\dots n) f(1) f(2) \cdots f(n)$
with $\bar{f}=1-f$. The solution of the equation $\psi_4 = K_4(4 \mu)
\psi_4$ gives the critical temperature for the onset of quartetting as 
a function of the chemical potential as shown in
Fig.~\ref{fig_tc-mue}, or the density as shown in Fig.~\ref{fig_tc-n}.
Within an estimate by using a variational calculation, the
transition to quartetting beats the transition to isospin singlet
pairing if the density is smaller than $0.03\fm^{-3}$,
see Fig.~\ref{fig_tc-n}.
\subsection{Gap equation and condensation energy for asymmetric
  nuclear matter}
For infinite nuclear matter, the gap energy at zero temperature as
well as at finite temperature has been investigated for pairing in the
different channels in dependence on nucleon density and isospin asymmetry.
In particular, it has been found that the gap energy in the isospin
singlet channel is strongly reduced for increasing asymmetry, and the
transition to superfluidity is possible only for asymmetry values
$\alpha=(n_n-n_p)/(n_n+n_p)\leq 0.35$ \cite{AFRS,SAL}. Also, the critical
temperature is strongly suppressed with increasing asymmetry as it can be
calculated from the gap energy as well as directly from the solution of
the BCS equation.

We give some relevant expressions for a fermion system interacting via 
a separable potential
\be
\label{int}
V_{\tau\tau'} = - \lambda \sum_{P,k,k'} w(k) w(k')\,\, a^\dagger_{\tau}(P/2+k)\,\,
a^\dagger_{\tau'}(P/2-k) \,\, a_{\tau'}(P/2-k') \,\, a_{\tau}(P/2+k')\,.
\ee
We use a Yamaguchi type of potential \cite{yama} with
$w(k) = (k^2 + \kappa^2)^{-1}$, and $\kappa=1.4488\fm^{-1}$.
The interaction strength in the $^1{\rm S}_0$ channel is only about
70 percent of the strength in the $^3{\rm S}_1$ channel which in its
original form \cite{yama} is chosen to reproduce
the deuteron binding energy as well as the low-energy behavior of
the free scattering phase shifts.
We perform an exploratory calculation and
consider the interaction strength as a parameter that will be adjusted below.
Furthermore, only zero angular momentum is considered.
Separable representations of more realistic interactions can be found
in the literature \cite{sep}.

The interaction is treated in mean-field (Hartree-Fock-Bogoliubov)
approximation, allowing for an isospin-singlet pair amplitude at zero
total momentum $P$. Diagonalizing $H^{\rm MF}-\mu_p N_p -\mu_n N_n$
using the Bogoliubov transformation, we obtain the gap equation 
\be
\label{gap}
\Delta(k) =  \lambda w(k) \sum_{k'} w(k') {\Delta(k') \over
  \sqrt{(\xi_{p}(k') + \xi_{n}(k'))^2 + 4 \Delta^2(k')}} 
\left[ 1-f(E^+_{k'})-f(E^-_{k'}) \right] 
\ee
with
\be
E^\pm_k = {1 \over 2} \left[\sqrt{(\xi_{p}(k) + \xi_{n}(k))^2 + 4
    \Delta^2(k)} \pm (\xi_{p}(k) - \xi_{n}(k))\right] 
\ee
and $\xi_{p}(k) = \epsilon_p(k) - \mu_p$,
$\xi_{n}(k) 
= \epsilon_n(k) - \mu_n$, $f(E) = ({\rm e}^{E/T}+1)^{-1}$. 
$ \epsilon_\tau(k)$ are the single-particle energies including the
shift due to a mean field as given above (\ref{eps}).
From the self-consistent solution of the 
gap equation, besides the trivial solution $\Delta(k) = 0$ also a
solution $\Delta(k) = g\,\, w(k)$ with a finite value of $g$ may occur.

The shift in the energy density due to the formation of a gap
(condensation energy density) follows as \cite{FW}
\bea
\label{conden}
&&\Delta{\cal E}(n_p,n_n) =
[{\cal E}_{\rm pair}(n_p,n_n)-{\cal E}_{\rm norm}(n_p, n_n)] = \nonumber\\
&&\frac{2}{V}\sum_k\Bigg\{ {1 \over 2}
\left[1-\frac{\xi_{p}(k) + \xi_{n}(k)}{\sqrt{(\xi_{p}(k) +
    \xi_{n}(k))^2 + 4 \Delta^2(k)}}\right] (\epsilon_p(k)
+ \epsilon_n(k))\nonumber\\
&&\qquad-{}{\Delta^2(k) \over  \sqrt{(\xi_{p}(k) +
    \xi_{n}(k))^2 + 4 \Delta^2(k)}}
  \left[ 1+f(E^+_{k})+f(E^-_{k}) \right]\nonumber\\
&&\qquad+{1 \over 2} f(E^+_{k}) \left[ (\epsilon_p(k)-
  \epsilon_n(k)) + 
(\epsilon_p(k)+ \epsilon_n(k)) {\xi_{p}(k) + \xi_{n}(k) \over
\sqrt{(\xi_{p}(k) + \xi_{n}(k))^2+ 4 \Delta^2(k)}}\right.\nonumber\\
&&\qquad\qquad\qquad\left. +{}4 {\Delta^2(k) \over
    \sqrt{(\xi_{p}(k) + 
    \xi_{n}(k))^2 + 4 \Delta^2(k)}} \right]\nonumber\\
&&\qquad+{1 \over 2}f(E^-_{k}) \left[(\epsilon_n(k)-
  \epsilon_p(k)) + 
(\epsilon_p(k)+ \epsilon_n(k)) {\xi_{p}(k) + \xi_{n}(k) \over
  \sqrt{(\xi_{p}(k) +  
    \xi_{n}(k))^2 + 4 \Delta^2(k)}}\right.\nonumber\\
&&\qquad\qquad\qquad\left. +{}4 {\Delta^2(k) \over
    \sqrt{(\xi_{p}(k) + 
    \xi_{n}(k))^2 + 4 \Delta^2(k)}} \right]\nonumber\\
&&\qquad-\epsilon_p(k) f( \epsilon_p(k)-\mu_p) -
\epsilon_n(k) f( 
\epsilon_n(k)-\mu_n)\Bigg\}\,.
\eea
The chemical potentials are given by the normalization to the
densities of the corresponding nucleons 
\be
\label{rho}
 {2 \over V} \sum_k f(
\epsilon_\tau(k)-\mu_\tau) = n_\tau \,\,.
\ee
As usual, it is assumed
that the normalization condition in the paired state gives
no essential change in the corresponding chemical potentials \cite{FW}.

\subsection{Finite nuclei density profiles}
\label{sec-dens}
For infinite nuclear matter, the energy density is calculated for
homogeneous densities $n_\tau$. In finite nuclei, the densities
$n_\tau (r)$ are depending on position $r$. Then also the quantities
considered above which are functions of the densities now are
parametrically depending on $r$.

If the density distribution of protons and neutrons is known, the
gain of the binding energy due to $np$ pairing (condensation energy) can be
estimated in LDA by the integral
\begin{equation}
\label{lda}
B_{ np} \approx 4\pi\int_0^\infty r^2 \Delta{\cal E}(n_p,n_n;r)
 dr\,\,.
\end{equation}

For nuclei with $N$ neutrons and $Z$ protons and $A=N+Z$,
we have to determine the density profiles 
of protons and neutrons. Exploratory calculations will be performed
taking the nucleon densities from a simple potential model, normalized 
to the corresponding numbers of protons or neutrons. 

The nucleons feel a mean field phenomenologically defined as
\be
\label{pot}
V_p (r) = V_p^{\rm nucl}(r) + V^{\rm coul}(r) , \qquad \qquad
V_n (r) = V_n^{\rm nucl}(r)\,\,. 
\ee
We adopt the Shlomo parameterization for the mean field \cite{Shlomo} :
\be
V_{\tau}^{\rm nucl} = \frac{V^0_{\tau}}{1+\exp((r-R)/d)} , \qquad \qquad
V^0_{\tau} = -V^0 + \tau V^{\rm sym} (N-Z)/A \,\,,
\ee
with the parameter values $V^0=54\MeV$, $V^{\rm sym}=33\MeV$, and
$d=0.7\fm$.
The radius of the nuclear potential is given by the implicit equation
\be
R = \frac{1.12 A^{1/3}+1.0}{ [1+(\pi d / R)^2]^{1/3}}\,\,.
\ee
For the Coulomb potential we use the charged sphere formula
\be
V^{\rm coul} =  \frac{Ze^2}{ 2R_c} \,\, \left[3-\frac{r^2}{R_c^2}
\right] \Theta(R_c-r)  + \frac{Ze^2}{r}   \Theta(r-R_c) \,\,,
\ee
where $e^2=1.44\MeV\fm$, $\Theta(x)$ denotes the step function. The
Coulomb radius is given by 
\be
R_c^2 =\frac{5}{3} <r^2>=C^2 \frac{1+ 10/3 (\pi z/C)^2+7/3 (\pi
z/C)^4 }{1+(\pi z/C)^2}  \,\,,
\ee
\be
C = \frac{1.12 A^{1/3}}{[1+(\pi z /C)^2]^{1/3}} \,\,,
\ee
where $<r^2>$ is m.s. radius from the nuclear charge density, $z=0.54\fm$.

Within the Thomas-Fermi approximation, at zero temperature the local
density is given by
\bea
n_{\tau}(r)\,&=&\,\frac{1}{3\pi^2} (k^F_{\tau})^3 \Theta
(\lambda_{\tau}-V_{\tau}(r))\,\, \nonumber\\
k^F_{\tau} \,&=&\, \left[\frac{2m}{\hbar^2}
  (\lambda_{\tau}-V_{\tau}(r)) \right]^{1/2}\,\,,
\eea
where $V_{\tau}(r)$ is the total potential (\ref{pot}). The chemical
potentials  $\lambda_{\tau}$ are determined by the constraints 
\be
Z = \int d^3r\,\, n_p (r) \,\,,\qquad \qquad \qquad
N = \int d^3r\,\, n_n (r) \,\,.  
\ee
Obviously this simple treatment gives only a first estimate of the
nucleonic densities. They are correctly normalized and account also
for the Coulomb repulsion, but can be improved by considering the
quasiparticle dispersion relation or shell effects in the nuclear
density. 
\section{Numerical results}
\subsection{Condensation energy}
\begin{figure}
\centerline{\psfig{figure=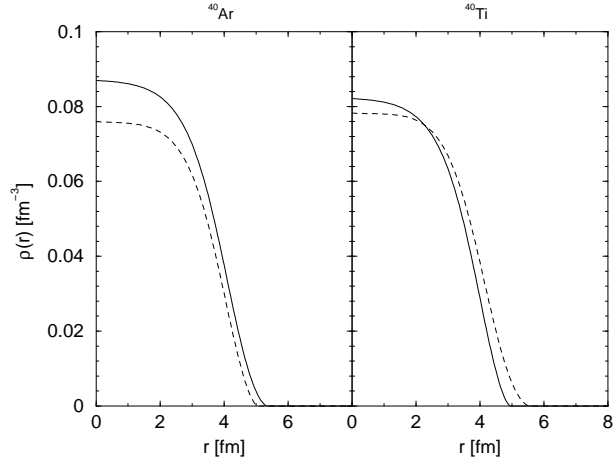,width=8cm}}
\caption{Density profiles of neutrons (solid lines) and protons
(dashed lines) calculated within the Thomas-Fermi approximation for
$^{40}{\rm Ar}$ and $^{40}{\rm Ti}$ at zero temperature.}
\label{fig_profiles}
\end{figure}

In the LDA, we proceed in the following way.
First, we calculate the density profiles $n_{\tau}(r)$ for protons 
and neutrons (see Sec.~\ref{sec-dens}).
As typical examples the density profiles for $^{40}{\rm Ar}$
and $^{40}{\rm Ti}$ are shown in Fig.~\ref{fig_profiles}.
Then, with the total local density
and the asymmetry $[n_n(r)-n_p(r)]/[n_n(r)+n_p(r)]$ as inputs the
local gap function $\Delta(k;r)$ is obtained
from the solution of the gap equation (\ref{gap}). This is performed
within a self-consistent Hartree-Fock scheme, which gives the local
single-particle energies $\epsilon_\tau(k;r)$ depending on the isospin
variable. Having the pairing gap at our
disposal the local condensation energy density (\ref{conden})
is determined. Finally,
integration over the whole nucleus (\ref{lda}) gives the contribution
to the binding energy due to $np$ pairing.

The coupling strength $\lambda$ of the separable interaction is
considered as a parameter which should be taken as a phenomenological 
quantity. To reproduce the average Wigner energy in the mass number region
 $20\leq A\leq 100$ ($2.62\MeV$), we have taken the 
value $\lambda=92.35\MeV\fm^3$ in the $^1{\rm S}_0$ channel and
$\lambda=131.50\MeV\fm^3$ in the $^3{\rm S}_1$ channel. Usually the
deuteron binding energy is used to adjust this parameter.
However, within our simple model we should take into
consideration that the interaction (\ref{int}) is an effective
description, simplifying different contributions as the different
channels, short distance repulsion, spin-orbit coupling etc.
Furthermore, some suppression of $T=0$ pairing in
symmetric nuclear matter due to medium polarization is expected.
The fact that the $np$ pair is bound (deuteron) whereas $nn$ and $pp$ pairs
are not is essentially due to the tensor force leading to the $d$-wave
component in the deuteron. Without this component the $pn$ interaction
in the $T=0$ channel would hardly be different from the $nn$ or $pp$
interactions. The fate of the tensor force in the nuclear medium is,
however, a much debated subject in nuclear physics and it is quite
possible that the tensor force is much more screened than the other
parts of the nuclear force \cite{Zamick}. In this sense the use of a bare
interaction in the $pn$ ($T=0$) channel may be more questionable than
it is in the $T=1$ channel.

\begin{figure}
\centerline{\psfig{figure=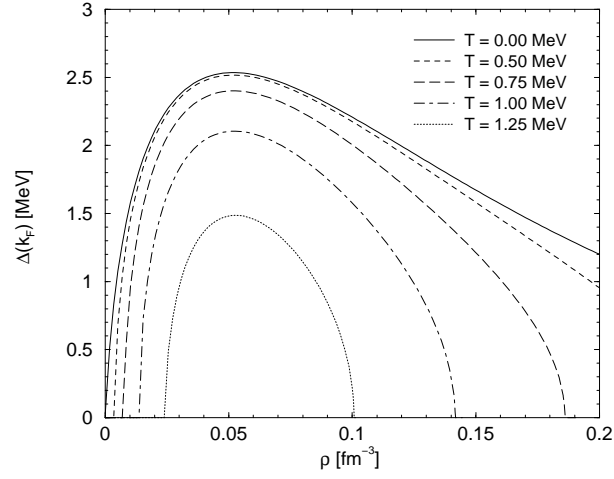,width=8cm}}
\caption{Proton-neutron pairing gap for asymmetric nuclear matter as a function of the total density for symmetric nuclear matter and different temperatures $T$.}
\label{fig_gap_t}
\end{figure}

\begin{figure}
\centerline{\psfig{figure=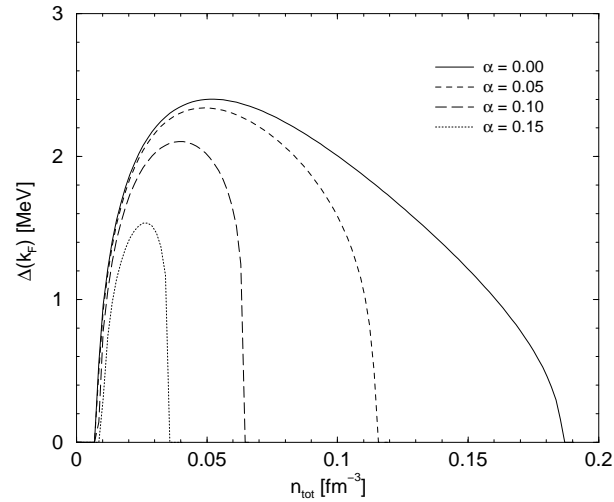,width=8cm}}
\caption{Proton-neutron pairing gap for asymmetric nuclear matter as a function of the total density at $T=0.75\MeV$ temperature and four values of the asymmetry parameter $\alpha$.}
\label{fig_gap_a}
\end{figure}

The isospin singlet gap for symmetric nuclear matter is shown in
Fig.~\ref{fig_gap_t} for different temperatures.
Below normal nuclear matter density,
which is of relevance here,
the difference between $T=0$ and $T=0.5\MeV$ temperature is small.
The dependence of the isospin singlet gap on the nuclear matter
density is shown in Fig.~\ref{fig_gap_a} for the temperature
$T=0.75\MeV$ and different asymmetries. 
The gap is strongly reduced for increasing asymmetries and
temperatures confirming the results obtained previously in
\cite{AFRS,SAL}.

\begin{figure}
\centerline{\psfig{figure=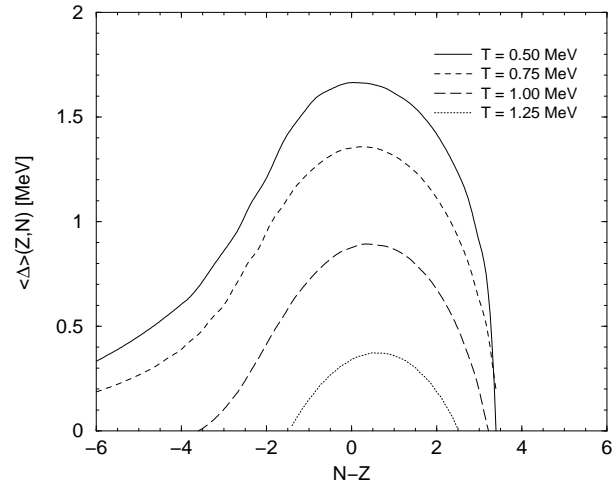,width=8cm}}
\caption{Average proton-neutron pairing gap of nuclei with a mass number $A=40$ as a function of the asymmetry $N-Z$ and three values of the temperature.}
\label{fig_avgap}
\end{figure}

For a nucleus with $A= 40$, the average gap on the Fermi level \cite{ST}
\be
\Delta = \sum_\tau \int d^3r \Delta(k_F(r);r) [n_\tau(r)]^{\frac{1}{3}}
\bigg/  \sum_\tau \int d^3r [n_\tau(r)]^{\frac{1}{3}} 
\ee
is shown as function of the asymmetry in Fig.~\ref{fig_avgap} for
different temperatures.

\begin{figure}
\centerline{\psfig{figure=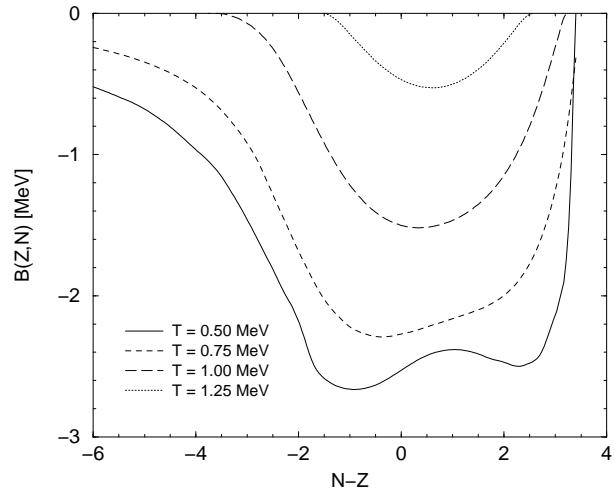,width=8cm}}
\caption{Condensation energy of nuclei with a mass number $A=40$ as a function of the asymmetry $N-Z$ and three values of the temperature.}
\label{fig_cond}
\end{figure}

The condensation energy as a function of asymmetry is shown in
Fig.~\ref{fig_cond} for different temperatures.
For the calculation, a fixed $A=40$ was assumed. Below $T=0.5\MeV$,
the dependence on temperature is negligible (see also
Fig.~\ref{fig_gap_t}), but becomes strong
for $T>1\MeV$. The influence of the Coulomb interaction taken into
account for the calculation of the density profiles
destroys the symmetry with respect to $N-Z=0$.
The Coulomb effect is to increase the overlap between neutron and proton
densities in the tail of the density profiles,
as show in Fig.~\ref{fig_profiles}. As a
consequence the pairing gap is slightly enhanced. Furthermore, the
steep decrease of the condensation energy near $N-Z=5$ is also
shown. This is in correspondence to the findings given in
Fig.~\ref{fig_bh-diff}.
\subsection{Quartetting}
The additional contribution due to quartetting seems to be high in the 
region of light nuclei, where the $\alpha$ cluster model is a good
approximation. The strong even-odd staggering is reduced at higher
masses.

To give an estimation of the effect of quartetting, the evaluation of 
the $np$ condensation energy (\ref{lda}) was repeated for symmetric
nuclei ($Z=N$), where the magnitude of the gap $\Delta(k;r)$ was
increased in the density region where quartetting can occur,
i.e. at densities below $0.03\fm^{-3}$. In detail, the gap was
increased by a factor, which was obtained from the ratio $T^c_4/T^c_t$ 
of the corresponding critical temperatures at given density $n$, as
shown in Fig.~\ref{fig_tc-n}.

Comparing with calculations neglecting quartetting, the gain of
binding energy due to quartetting has been evaluated for different
nuclei with $Z=N$. Exploratory calculations for nuclei of medium size
($A \approx 100$) show that the contribution due to
quartetting is almost zero but may become large for small
$A$. For instance, the calculation for $^{12}$C and $^{16}$O give an
additional contribution to the condensation energy due to quartetting
of 10.3 and 9.6 percent, respectively, if compared with isospin
singlet pairing.
\subsection{Excited Nuclei}
After discussing the contribution of $pn$-pairing to the nucleon
binding energy, comparing with the pairing energy in asymmetric
nuclear matter, it is of interest to discuss also the effect of
excitations. In nuclear matter, excitations are well understood in the
context of finite temperatures. It is expected that the effects of
condensates are decreasing with increasing excitation.

\begin{figure}
\centerline{\psfig{figure=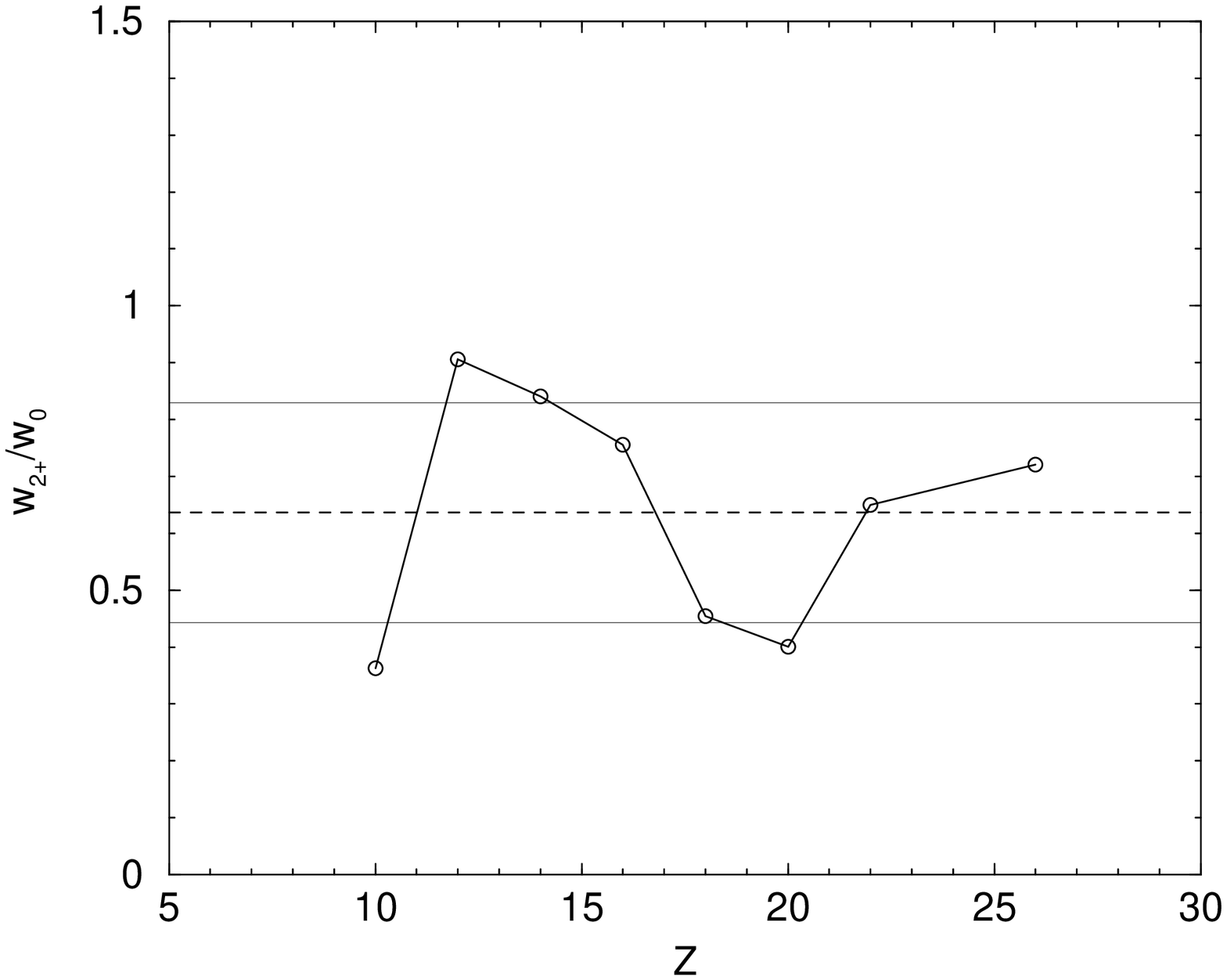,width=8cm}}
\caption{For the filter $w(Z,Z)$ the ratio $w_{2^+}/w_0$ of $2^+$ excited nuclei over ground state nuclei is given for even proton and neutron number. The average value is indicated by the dashed lines as well as the statistical error bars by the thin lines.}
\label{fig_excited}
\end{figure}

To investigate the effect of excitations on the formation of
condensates in finite nuclei, we analyse the $2^+$ excitations of
even-even nuclei. In particular, we used the $w$-filter for the analysis
and compared $w_{2^+}(Z,Z)$ for the excited nuclei with $w_0(Z,Z)$ for
the ground state nuclei,
see Fig.~\ref{fig_excited}. We determined the mean value of the
ratio from $Z=10$ to $Z=26$ and obtained $w_{2^+}(Z,Z)/w_0(Z,Z)=0.64\pm0.19$.

This result can be compared with the influence of finite temperature
on the pairing in nuclear matter. The average 2$^+$ excitation energy
of even-even nuclei, taken for the interval $10\leq Z\leq 30$, is
$1750\keV$. Performing a finite temperature Thomas-Fermi calculation
for a nucleus of medium proton number $Z=20$,
this excitation energy would correspond to a
temperature of about $1\MeV$.

As shown in Fig.~\ref{fig_cond}, similar to the decrease of the gap we observe
also a decrease of the value $b_0(Z)$ with increasing temperature. The 
result of the calculation is in agreement with the empirical value
given above.
\section{Conclusions}
In this work we have shown that isospin-singlet pairing and $\alpha$ like
quartetting may contribute to the binding energy of nuclei very close to
the symmetry line $N=Z$. These contributions are relatively large for
smaller nuclei ($Z<20$). For medium mass
nuclei neutron-proton pairing in the isoscalar channel
disappears already for $|N-Z|=4$. This stems from the very rapid decrease
of the isoscalar pairing as a function of the unbalance in the Fermi
energies of protons and neutrons.
These facts can explain the origin of the Wigner term in the mass formula as
well as the empirically determined $\Delta B(Z,N)$ or $b_{Z-N}(Z)$,
respectively (see Eqs.~(\ref{BW}), (\ref{bi_even-odd})).
An enhancement of the isoscalar pairing
contribution to the binding energy is obtained if in addition quartetting
is taken into account. An interesting effect is that the reduction of the
condensation energy with increasing excitation of the nuclei seems to be
in agreement with empirical data.

Our calculations are exploratory in the sense that they were performed
in the rather crude LDA approach. However, please notice that the LDA has
yielded in the past quite reasonable results on the average, i.e. for a gap
averaged over the shell effects \cite{ST}. Therefore, we think that our results
give a quite reliable first orientation of the effect.
The approach should be improved in several respects. First a more realistic
force should be employed. Second shell effects must properly be included.
Eventually, effects of number projection, even-odd staggering and pair
fluctuations should also be investigated.
Such studies shall be performed in the future.
It is the hope that isoscalar pairing and quartetting will give us precious
hints on the effective neutron-proton interaction
in a nuclear medium as well as very interesting clustering and condensation
phenomena in nuclei.
\section*{Acknowledgments}
We grateful acknowledge fruitful discussions with K. Langanke,
E. Moya de Guerra, A. Poves, and P. Ring.

\newpage
\newpage
\newpage
\newpage
\newpage
\newpage

\begin{thebibliography}{99}
\bibitem[*]{isn}
Permanent Address: Institut de Sciences Nucl\'{e}aires,
Universit\'{e} Joseph Fourier, CNRS-IN2P3
53, Avenue des Martyrs, F-38026 Grenoble Cedex, France

\bibitem{shell-model}
M. G. Mayer and J. H. D. Jensen,
{\it Elementary Theory of Nuclear Shell Structure},
(Wiley, New York, 1955).

\bibitem{BM}
A. Bohr and B. R. Mottelson
{\it Nuclear Structure}
(Benjamin, N.Y., 1969), Vol. I;
P. Ring and P. Schuck,
{\it The Nuclear Many-Body Problem}
(Springer, N.Y., 1980).

\bibitem{Arne}
T. Alm, G. R\"opke, A. Schnell, N. H. Kwong, and H. S. K\"ohler,
Phys. Rev. C {\bf 53}, 2181 (1996);
A. Schnell, T. Alm, and G. R\"opke,
Phys. Lett. B {\bf 387}, 443 (1996).

\bibitem{Wim}
W. H. Dickhoff and H. M\"uther,
Rep. Prog. Phys. {\bf 11}, 1947 (1992);
B. E. Vonderfecht, W. H. Dickhoff, A. Polls, and A. Ramos,
Nucl. Phys. {\bf A555}, 1 (1993).

\bibitem{RMS82}
G. R\"opke, L. M\"unchow, and H. Schulz,
Nucl. Phys. {\bf A379}, 536 (1982);
Phys. Lett. B {\bf 110}, 21 (1982);
G. R\"opke, M. Schmidt, L. M\"unchow, and H. Schulz,
Nucl. Phys. {\bf A399}, 587 (1982).

\bibitem{SRS}
M. Schmidt, G. R\"opke, and H. Schulz,
Ann. Phys. (NY) {\bf 202}, 57 (1990).

\bibitem{Nazarew}
J. Dobaczewski, W. Nazarewicz, T. R. Werner, J. F. Berger, C. R. Chinn,
and J. Decharg\'{e},
Phys. Rev. C {\bf 53}, 2809 (1996).

\bibitem{ShapT}
S. L. Shapiro and S. A. Teukolsky,
{\it Black Holes, White Dwarfs and Neutron Stars: The Physics of Compact Objects}
(Wiley, N.Y., 1983);
D. Pines, R. Tamagaki, and S. Tsurata (eds.),
{\it Neutron Stars}
(Addison-Wesley, N.Y., 1992).

\bibitem{BLS}
M. Baldo, U. Lombardo, and P. Schuck,
Phys. Rev. C {\bf 52}, 975 (1995).

\bibitem{ZPhys95}
H. Stein, A. Schnell, T. Alm, and G. R\"opke,
Z. Phys. A {\bf 351}, 295 (1995).

\bibitem{goodman}
A. L. Goodman,
Nucl. Phys. {\bf A352}, 30 and 45 (1981);
Nucl. Phys. {\bf A369}, 365 (1981).

\bibitem{goodman2}
A. L. Goodman,
Phys. Rev. C {\bf 58}, 3051 (1998);
Phys. Rev. C {\bf 60}, 014311 (1999).

\bibitem{Marumori}
T. Marumori and K. Suzuki,
Nucl. Phys. {\bf A106}, 610 (1968).

\bibitem{Cau81}
M. Cauvin, V. Gillet, F. Soulmagnon, and M. Damos,
Nucl. Phys. {\bf A361}, 192 (1981);
Y. K. Gambhir, P. Ring, and P. Schuck,
Phys. Rev. Lett. {\bf 51}, 1235 (1983);
K. Varga, R. G. Lovas, and R. J. Liotta,
Phys. Rev. Lett. {\bf 69}, 37 (1992);
F. Aldabe, G. G. Dussel, and H. M. Sofia,
Phys. Rev. C {\bf 50}, 1518 (1994).
G. G. Dussel, R. J. Liotta, and R. P. J. Perazzo,
Nucl. Phys. {\bf A388}, 606 (1982);
F. Catara, A. Insolia, and U. Lombardo,
Nucl. Phys. {\bf A261}, 282 (1976).

\bibitem{lett}
G. R\"opke, A. Schnell, P. Schuck, and P. Nozi\`eres,
Phys. Rev. Lett. {\bf 80}, 3177 (1998).

\bibitem{LDA}
R. O. Jones and O. Gunnarson,
Rev. Mod. Phys. {\bf 61}, 689 (1989);
R. M. Dreizler and E. K. U. Gross,
{\it Density Functional Theory}
(Springer, Berlin, 1990);
H. Eschrig,
{\it The Fundamentals of Density Functional Theory}
(Teubner, Leipzig, 1996).

\bibitem{ST}
H. Kucharek, P. Ring, P. Schuck, and R. Bengtson,
Phys. Lett. B {\bf 216}, 249 (1989);
P. Schuck and K. Taruishi, Phys. Lett. B {\bf 385}, 12 (1996).

\bibitem{Satula97}
W. Satula and R. Wyss,
Phys. Lett B {\bf 393}, 1 (1997).

\bibitem{Satula97a}
W. Satula, D. J. Dean, J. Gary, S. Mizutori, and W. Nazarewicz,
Phys. Lett B {\bf 407}, 103 (1997). 

\bibitem{Poves98}
A. Poves and G. Martinez-Pinedo,
Phys. Lett. B {\bf 436}, 19 (1998);
G. Martinez-Pinedo, K. Langanke, P. Vogel,
Nucl. Phys. {\bf A651}, 379 (1999).

\bibitem{Engellanganke}
J Engel, K. Langanke, and P. Vogel,
Phys. Lett. B {\bf 398}, 211 (1996);
Phys. Lett. B {\bf 429}, 215 (1998).

\bibitem{Langanke}
P. B. Radha, D. J. Dean, S. E. Koonin, and K. Langanke,
Phys. Rev. C {\bf 56}, 3079 (1997);
K. Langanke,
Nucl. Phys. A {\bf 630}, 368c (1998).

\bibitem{Frauendorf}
S. G. Frauendorf and J. A. Sheikh,
Phys. Rev. C {\bf 59}, 1400 (1999).

\bibitem{Civitarese}
O. Civitarese, M. Reboiro, and P. Vogel,
Phys. Rev. C {\bf 56}, 1840 (1997).

\bibitem{Kaneko}
K. Kaneko and M. Hasegawa,
nucl-th/9906001 in the xxx.lanl.gov archive.

\bibitem{AFRS}
T. Alm, G. R\"opke, M. Schmidt,
Z. Phys. A {\bf 337}, 355 (1990);
T. Alm, B. L. Friman, G. R\"opke, and H. Schulz,
Nucl. Phys. {\bf A551}, 45 (1993);
B. E. Vonderfecht, C. C. Gearhart, W. H. Dickhoff, A. Polls, and A. Ramos,
Phys. Lett. B {\bf 253}, 1 (1992);
M. Baldo, I. Bombaci, and U. Lombardo,
Phys. Lett. B {\bf 283}, 8 (1993).

\bibitem{SAL}
A. Sedrakian, T. Alm, and U. Lombardo,
Phys. Rev. C {\bf 55}, R582 (1997).  

\bibitem{RSSnashville}
G. R\"opke, A. Schnell, and P. Schuck,
proceedings of CMT22, Nashville, 1998

\bibitem{erice}
G. R\"opke and A. Schnell,
Prog. Part. Nucl. Phys. {\bf 42}, 53 (1999).

\bibitem{dukelski}
G. R\"opke,
Ann. Pysik {\bf 3}, 145 (1994);
J. Dukelsky, G. R\"opke, and P. Schuck,
Nucl. Phys. {\bf A 628}, 17 (1998).

\bibitem{yama}
Y. Yamaguchi,
Phys. Rev. {\bf 95}, 1628 (1954).

\bibitem{sep}
J. Haidenbauer and W. Plessas,
Phys. Rev. {\bf C 30}, 1822 (1984);
C. Brandstaetter, Diploma Thesis, Univ. Graz, 1993 (unpublished);
W. Plessas et al., Few-Body Systems Suppl. {\bf 7}, 251 (1994);
H.-P. Kotz et al., in Few-Body Problems in Physics (AIP Conference
Proceedings, Vol. {\bf 334}), ed. by F. Gross, New York, 1995, p. 482.

\bibitem{FW}
A. L. Fetter and J. D. Walecka,
{\it Quantum Theory of Many-Particle Systems}
(McGraw-Hill, N.Y., 1971).

\bibitem{Shlomo}
S. Shlomo,
Nucl. Phys. {\bf A539}, 17 (1992).

\bibitem{Zamick}
M. S. Fayache and L. Zamick,
Phys. Rev. C {\bf 51}, 160 (1995);
L. Zamick, D. C. Zheng, and M. S. Fayache,
Phys. Rev. C {\bf 51}, 1253 (1995);
U. Lombardo,
{\it Superfluidity in Nuclear Matter}, In:
{\it Nuclear Methods and Nuclear Equation of State},
cap.IX, Ed. M. Baldo, (World Scientific, Singapore, 1999).
\end{thebibliography}
\end{document}